\PassOptionsToPackage{table,dvipsnames}{xcolor}   
\documentclass[conference, 9pt]{IEEEtran}
\IEEEoverridecommandlockouts
\usepackage{amsmath,amsfonts}
\usepackage{algorithmic}
\usepackage{algorithm}
\usepackage{array}
\usepackage{textcomp}
\usepackage{stfloats}
\usepackage{url}
\usepackage{verbatim}
\usepackage{graphicx}
\usepackage{cite}

\usepackage{balance}
\usepackage{amsmath,amssymb,amsfonts}

\usepackage[dvipsnames]{xcolor}

\usepackage{amssymb}
\usepackage{amsthm}
\usepackage{bm}
\usepackage{booktabs}
\usepackage{color}
\usepackage{cite}             
\usepackage{comment}          
\usepackage{enumitem}
\usepackage{etoolbox}         
\usepackage{graphicx}         
\usepackage{mathtools}
\usepackage{multirow}
\usepackage{pgfplots}
\usepackage{pgfplotstable}
\usepackage{setspace}
\usepackage{stfloats}
\usepackage{caption}
\usepackage{subcaption}
\usepackage{tabularx}
\usepackage{threeparttable}
\usepackage{tikz}
\usepackage{url}
\usepackage{dsfont}
\usepackage{float}

\pgfplotsset{compat=newest}

\newcolumntype{Y}{>{\centering\arraybackslash}X}

\theoremstyle{definition}

\renewrobustcmd{\bfseries}{\fontseries{b}\selectfont}
\renewrobustcmd{\boldmath}{}
\newrobustcmd{\B}{\bfseries}

\usetikzlibrary{decorations.pathreplacing,calc}

\usepackage{colortbl}     
\usepackage{xcolor}       
\usepackage[
  colorlinks,
  linkcolor=blue,
  citecolor=magenta,
  filecolor=Mulberry,
  urlcolor=NavyBlue]{hyperref}

\usepackage{soul}
\usepackage{graphicx}
\graphicspath{{./figs/}}
\usepackage{booktabs}
\usepackage{multirow}
\usepackage[square,sort,numbers,compress]{natbib}

\newcommand{\todo}[1]{\textcolor{blue}{\textbf{[TODO: #1]}}}
\newcommand{\new}[1]{\textcolor{red}{#1}}
\newcommand{\newnew}[1]{\textcolor{blue}{#1}}
\definecolor{lightgreen}{RGB}{144,238,144}
\definecolor{lightblue}{RGB}{173,216,230}

\newcommand{\name}{\texttt{PPAAS}\space}

\begin{document}
\title{PPAAS: \underline{P}VT and \underline{P}areto \underline{A}ware \underline{A}nalog \underline{S}izing via Goal-conditioned Reinforcement Learning \\
}


\author{
Seunggeun Kim$^{1}$, Ziyi Wang$^{1,2\dagger}$\thanks{$\dagger$: Work done while the author was a visiting student at UT Austin.}, Sungyoung Lee$^{1}$,  Youngmin Oh$^{3}$, Hanqing Zhu$^{1}$, Doyun Kim$^{3}$, David Z. Pan$^{1}$ \\
$^{1}$\textit{The University of Texas at Austin}, \quad $^{2}$\textit{Chinese University of Hong Kong}, 
\quad 
$^{3}$\textit{Samsung Electronics} \\
}

\maketitle
\begin{abstract}
Device sizing is a critical yet challenging step in analog and mixed-signal circuit design, requiring careful optimization to meet diverse performance specifications.
This challenge is further amplified under process, voltage, and temperature (PVT) variations, which cause circuit behavior to shift across different corners.
While reinforcement learning (RL) has shown promise in automating sizing for fixed targets, training a generalized policy that can adapt to a wide range of design specifications under PVT variations requires much more training samples and resources.
To address these challenges, we propose a \textbf{Goal-conditioned RL framework} that enables efficient policy training for analog device sizing across PVT corners, with strong generalization capability.
To improve sample efficiency, we introduce Pareto-front Dominance Goal Sampling, which constructs an automatic curriculum by sampling goals from the Pareto frontier of previously achieved goals. This strategy is further enhanced by integrating Conservative Hindsight Experience Replay to stabilize training and accelerate convergence.
To reduce simulation overhead, our framework incorporates a Skip-on-Fail simulation strategy.
Experiments on benchmark circuits demonstrate $\sim$1.6$\times$ improvement in sample efficiency and $\sim$4.1$\times$ improvement in simulation efficiency compared to existing sizing methods.
Code and benchmarks are publicly available \href{https://github.com/SeunggeunKimkr/PPAAS}{HERE}.
\end{abstract}

 \section{Introduction}


Analog and mixed-signal (AMS) circuit design is a critical yet highly challenging endeavor. 
\textit{Device sizing} serves as the primary method for optimizing AMS circuits to meet target specifications across various competing performance metrics. 
However, this process involves searching a vast design space that includes parameters such as transistor widths and lengths, as well as the resistance and capacitance of passive components. 
Additionally, non-idealities like PVT variations require the sized analog circuits to maintain their performance under diverse conditions. 
These complexities incur significant human expert involvement and excessive simulation feedback, leading to prolonged design times.

\begin{figure}
    \centering
    \begin{subfigure}[t]{\linewidth}
        \centering
        \includegraphics[width=\linewidth]{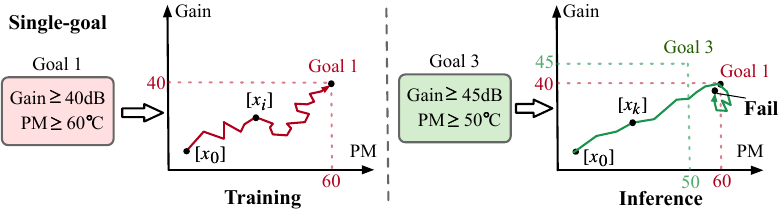}
        \caption{Single-goal RL fails to generalize to unseen goals}
    \end{subfigure}
    
    \vspace{0.5em}

    \begin{subfigure}[t]{\linewidth}
        \centering
        \includegraphics[width=\linewidth]{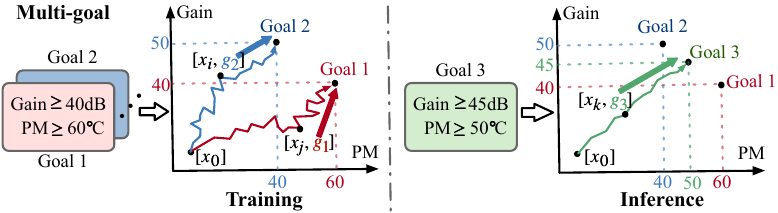}
        \caption{Goal-conditioned RL succeeds in achieving unseen goals}
    \end{subfigure}

    \caption{Key differences between Single-goal RL and GCRL. GCRL is designed to learn policies that generalize across goals. $x_i$ denotes the design parameters (state), and trajectories vary with different goals, starting from a shared initial state $x_0$.}
    \label{fig:intro}
    \vspace{-20pt}
\end{figure}

To accelerate the AMS circuit design process, various automation algorithms and methodologies for device sizing have been developed.
Early approaches, such as optimization-based techniques using Genetic Algorithms~\cite{Analog_TCAS_GA} and Bayesian Optimization~\cite{bo_analog}, lacked the capability to deal with complex design spaces.
In contrast, recent sizing approaches based on deep reinforcement learning (RL) have shown promising progress in handling complex scenarios, including \textit{larger circuits}~\cite{complex_circuit, complex_circuit2}, \textit{local mismatches}~\cite{parastic, cronus, pvtsizing}, and \textit{global variations}~\cite{Yang_2021,shi2022robustanalogfastvariationawareanalog,cao2024roseoptrobustefficientanalog, pvtsizing}.
These RL-based techniques fall into two main categories:
single-goal approaches~\cite{complex_circuit, complex_circuit2, parastic, Analog_DAC20_Wang, DNNopt_Budak, gmid, Yang_2021, shi2022robustanalogfastvariationawareanalog, pvtsizing} and multi-goal approaches~\cite{autockt, cao2024roseoptrobustefficientanalog, cronus}. Single-goal RL approaches are trained for a single target specification; they therefore lack \textit{generalizability} and require retraining when addressing new specifications.
On the other hand, \textcolor{black}{multi-goal approaches aim to handle multiple target goals simultaneously, and the goal-conditioned reinforcement learning (GCRL) approaches \cite{autockt, cao2024roseoptrobustefficientanalog, cronus} are most widely used among them.}
As shown in Fig.~\ref{fig:intro}, in the GCRL formulation, a distinct goal is assigned to each episode to handle multiple target goals using a single neural network, enabling the agent to learn a policy that generalizes across goals.
GCRL approaches indeed offer the ability to manage unseen goals at inference time without retraining.
However, they typically require \textit{significantly more training steps} than single-task methods, as they aim to solve a more generalized and complex problem.
Furthermore, goal assignment in each episode is crucial, as random selection may cause the agent to waste learning opportunities on goals that are either too easy or too difficult.


    

Incorporating PVT variations into RL-based device sizing further introduces significant complexity through two primary mechanisms.
\textit{First}, the computational burden grows prohibitively expensive due to linear scaling with PVT corner count, particularly constraining the parallelism of on-policy RL algorithms like Proximal Policy Optimization (PPO) \cite{ppo}. 
\textit{Second}, different variations can conflict with one another, complicating the circuit optimization problem and thus requiring more samples to train the agent.
Existing PVT-aware sizing methods~\cite{Yang_2021,shi2022robustanalogfastvariationawareanalog,cao2024roseoptrobustefficientanalog,pvtsizing} attempt to balance training efficiency and simulation cost during the training process.
However, these approaches either incur high simulation costs to ensure stable training across all PVT corners~\cite{cao2024roseoptrobustefficientanalog}, or risk suboptimal solutions due to degraded sample quality—particularly because k-means clustering does not reliably identify critical corners in multi-goal settings~\cite{ shi2022robustanalogfastvariationawareanalog,pvtsizing}.
This dilemma underscores the need for more adaptive multi-goal optimization frameworks that can produce high-quality solutions while managing the high training expenses associated with multi-PVT environments.

To address this dilemma, we propose an efficient GCRL framework named \name, short for PVT and Pareto-Aware Analog Sizing via GCRL.
Our method maintains high sample efficiency in multi-PVT environments, even when training samples are partially degraded due to the Skip-on-Fail strategy, which skips full-corner simulations when nominal-corner simulations fail to meet target specifications.
Our key contributions are summarized as follows:

\begin{itemize}
\item We propose a Pareto-Dominant Goal Sampling (PGDS) strategy that constructs an automatic curriculum by selecting hard enough goals from the Pareto frontier of previously achieved goals, thereby improving sample efficiency.
    
\item We design a novel goal representation tailored for multi-PVT environments, integrating it with Conservative Hindsight Experience Replay and a PVT-aware hierarchical reward formulation.
    
\item Experimental results show that \name achieves $\sim$1.6$\times$ improvement in sample efficiency and $\sim$4.1$\times$ improvement in simulation efficiency as compared to existing methods, demonstrating superior solution quality and efficiency.

\end{itemize}

The remainder of this paper is organized as follows: Section~\ref{background} provides preliminaries on the AMS circuit design, goal-conditioned RL, and prior approaches to PVT-aware analog sizing. In Section~\ref{method}, we present our GCRL framework and elaborate on our main contributions. Section~\ref{experiments} presents the ablation study and performance comparisons with prior work, followed by conclusions in Section~\ref{conclusion}.
\section{Preliminaries}
\label{background}
\begin{figure*}[t]
    \centering
\includegraphics[width=1.0\textwidth]{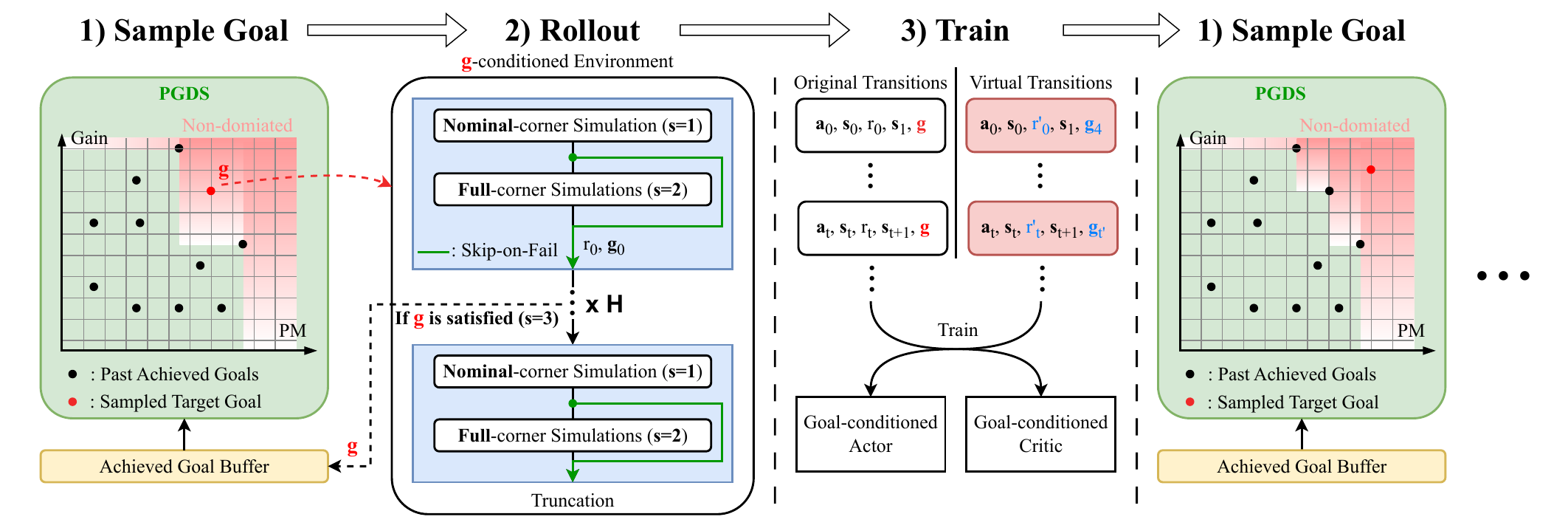}
    \caption{Overview of the proposed training process: It iteratively samples a goal using Pareto-Dominant Goal Sampling, rolls out an episode with the sampled goal, and updates the goal-conditioned actor and critic networks.}
    \label{fig:overview}
    \vspace{-10pt}
\end{figure*}

\subsection{Analog and Mixed-signal Circuit Design}
In AMS circuit design, achieving robust performance requires addressing both local mismatches and global variations.
Specifically, global variations refer to systematic fluctuations in PVT that affect all devices across a chip or wafer similarly, often caused by inconsistencies in manufacturing conditions or environmental factors.
Among PVT factors, process variations arise from manufacturing inconsistencies and are typically characterized using corner models, including fast-fast (FF), slow-slow (SS), fast-slow (FS), slow-fast (SF), and typical-typical (TT), with TT representing the nominal condition. Voltage variations occur due to fluctuations in supply voltage, e.g., $\pm10\%$ deviations from the nominal $V_{DD}$.
Temperature variations can range from $-40^{\circ}C$ to $125^{\circ}C$, with a standard operating temperature of $27^{\circ}C$~\cite{circuit_pvt_1, circuit_pvt_2}.
The interplay of these PVT variations can lead to substantial deviations from the circuit’s nominal design expectations, demanding that all specifications be rigorously satisfied across every corner and condition.
Furthermore, robust circuit design aims to maintain minimal deviation from nominal-corner specifications under all PVT variations.

This multi-corner verification process is computationally intensive in principle, as it demands additional simulation resources to evaluate and optimize circuit performance under each scenario.
Traditional design strategies address this by initially designing the circuit with performance margin at nominal conditions and then fine-tuning device parameters through simulations across all extreme PVT corners to minimize the cost of multi-corner analysis.
This manual approach is time-consuming and relies on heuristic tuning of performance margins, underscoring the need for automated methods capable of minimizing resources to handle PVT variations.

\subsection{Goal-conditioned Reinforcement Learning}
Rather than training a separate policy for each new goal, GCRL learns a single, universal policy capable of solving multiple goals~\cite{Schaul2015UniversalVF}.
Concretely, at the start of each episode, a goal $\mathbf{g} \in \mathcal{G}$ is sampled and remains fixed throughout the episode. 
Let $\mathcal{S}$ and $\mathcal{A}$ denote the state and action spaces of a standard RL setting, and $\mathcal{G}$ be the goal space. 
In GCRL, the goal-conditioned policy $\boldsymbol{\pi}(\mathbf{a}_t | \mathbf{s}_t, \mathbf{g})$ is trained to maximize the expected discounted return:
\begin{equation}
    \mathbb{E}_{\mathbf{s}_{0:H-1}, \mathbf{a}_{0:H-1}, \mathbf{g}} \left[ \sum_{t=0}^{H-1} \gamma^t r_t(\mathbf{s}_t, \mathbf{a}_t, \mathbf{g}) \right],
\end{equation}
where $r_t: \mathcal{S} \times \mathcal{A} \times \mathcal{G} \rightarrow \mathbb{R}$ is the reward function, and $H$ is the horizon length.
In practice, GCRL often operates in environments with sparse rewards, such as binary or discretized feedback, making the learning signal especially challenging.

The GCRL problem can be framed as a standard RL problem whose state space is augmented to $\mathcal{S} \times \mathcal{G}$, while the action space remains $\mathcal{A}$. 
Accordingly, we can learn a goal‐conditioned policy $\boldsymbol{\pi}: \mathcal{S} \times \mathcal{G} \rightarrow \mathcal{A}$ and a corresponding Q-function $Q: \mathcal{S} \times \mathcal{G} \times \mathcal{A} \rightarrow \mathbb{R}$ using standard RL algorithms, as demonstrated in prior works~\cite{Schaul2015UniversalVF, andrychowicz2018hindsightexperiencereplay, settaluri2020autocktdeepreinforcementlearning, cao2024roseoptrobustefficientanalog}.

This approach is particularly effective in environments with diverse but related goals, such as AMS circuit design, where target performance depends on the specific task and conditions~\cite{autockt, cronus, cao2024roseoptrobustefficientanalog}.
Once trained, GCRL enables zero-shot generalization to unseen goals, requiring as few as 30–60 simulations to solve new target specifications.
However, this flexibility comes at the cost of training complexity: GCRL must learn a generalized policy over a wide goal space, which significantly increases data requirements and makes training more challenging compared to single-goal RL.

To address these challenges in sparse reward settings, several augmentation techniques have been proposed within the GCRL framework. One prominent approach is Hindsight Experience Replay (HER)~\cite{andrychowicz2018hindsightexperiencereplay}, which enhances learning by manipulating the replay buffer in off-policy algorithms. 
Specifically, it assigns new goals to unsuccessful episodes based on the outcomes achieved, allowing the agent to learn from all interactions with the environment. 
This reinterpretation allows the agent to extract meaningful learning signals even from unsuccessful trials, effectively utilizing all collected experiences to improve sample efficiency.
Another line of work focuses on curriculum learning, which aims to shape the goal distribution $p(\mathbf{g})$ to guide the agent through a more structured learning progression~\cite{curriculum_learning, automatic_curriculum}.
Rather than sampling goals uniformly, curriculum methods adaptively modify $p(\mathbf{g})$ based on the agent’s current policy, encouraging training on goals that are neither too easy nor too hard.

However, the benefits of HER and curriculum learning are mostly pronounced in sparse-reward environments, where feedback is infrequent and uninformative for most goals. In contrast, when rewards are dense, meaning every interaction provides rich feedback about the environment dynamics, these techniques become less critical. In fact, applying HER or curriculum learning naively in dense-reward settings can lead to reduced sample efficiency or even slight performance degradation, as their assumptions no longer hold and may interfere with learning from already informative signals~\cite{quasi_reward}.

\subsection{Existing PVT-aware Analog Sizing Methods}
\label{section: related works}
Existing PVT-aware analog sizing methods can be broadly categorized based on whether they adopt a single-goal or GCRL framework.

RobustAnalog and PVTSizing~\cite{shi2022robustanalogfastvariationawareanalog, pvtsizing} represent approaches based on single-goal RL. These methods reduce simulation cost by identifying a subset of the most critical PVT corners using K-means clustering and apply multi-task RL over this reduced set. While effective in static settings, this strategy assumes that the identified corners remain consistently critical, which lowers sample quality in a GCRL setting where target specifications vary dynamically across episodes. As a result, these methods may converge to suboptimal solutions or fail to generalize when extended to multi-goal objective.

On the other hand, RoSE-Opt\cite{cao2024roseoptrobustefficientanalog} is explicitly designed within the GCRL framework. It trains a universal policy using the PPO algorithm\cite{ppo}, conditioning both the policy and value functions on varying goals while incorporating all PVT corners during training. Additionally, it leverages BO to initialize the RL agent, aiming to accelerate training by finding near-optimal initial state to start with. 

Nevertheless, as RoSE-Opt performs full corner simulations at every step, the simulation cost becomes prohibitively high unless a parallel simulator such as Cadence Spectre APS is used to amortize the workload.
Without such a simulator, the available resources are heavily consumed by multi-corner simulations, leaving limited capacity for parallel rollout. As PPO is an on-policy algorithm that relies on batched rollout workers to collect fresh trajectories for each update, it becomes a less attractive choice under such resource-constrained conditions.

These limitations call for a GCRL-based approach that improves time efficiency by jointly optimizing sample efficiency and the number of corners evaluated per sample under resource constraints.

\section{PPAAS Algorithms and Methods}\label{method}


\subsection{Skip-on-Fail Simulation Framework}
\label{SoF}

Before introducing the GCRL framework, inspired by the expert design process, we first highlight a simple yet effective hierarchical simulation strategy, termed the Skip-on-Fail (SoF) approach.
In the first stage ($s\!=\!1$), simulations are performed only under the nominal corner.
If the agent fails to meet the target specifications at this stage, the second stage ($s\!=\!2$)—comprising full corner simulations—is skipped, and the measured metrics are directly employed, thereby reducing redundant simulations.
Additional simulations across other corner conditions are conducted only when the nominal corner metrics exceed the target goal specifications. Note that the stage indicator $s$ is conceptually distinct from the state variable $\mathbf{s}_t$.
While similar techniques have been explored in single-goal settings~\cite{Yang_2021, pvtsizing}, we extend this strategy to the multi-goal setting, by conditionally pruning all corners except the nominal corner, without explicitly identifying critical corners.
Although the SoF method is conceptually simple and effectively reduces training time by lowering simulation costs, we emphasize that it does not inherently guarantee high sample efficiency, as this reduction comes at the expense of sample quality.
To address this limitation, we propose solutions detailed in Sections~\ref{pgds}, \ref{reward}, and~\ref{CHER}.
We further note that $s\!=\!3$ is assigned when the metrics measured in the second stage satisfy the target specifications.
A concise visual representation of the SoF rollout workflow is provided in step 2 of Fig.~\ref{fig:overview}.

\subsection{Goal-conditioned RL Setup and Notations}

In this subsection, we formally define the essential components and foundational structure of our GCRL framework.
It is modeled as a finite-horizon Markov Decision Process, represented by the tuple $(\mathcal{S}, \mathcal{A}, \mathcal{G}, \mathbf{f}, \mathbf{T}, \mathbf{s}_0, r, H)$.
Here, the state space is denoted by $\mathcal{S} \subseteq \mathbb{R}^L$ and the action space by $\mathcal{A} \subseteq \mathbb{R}^L$, where $L$ indicates the number of parameters to be optimized.
The goal space $\mathcal{G} \subseteq \mathbb{R}^M \times \{0,1\}^2$ includes continuous specification values and binary success indicators, with $M$ being the number of specifications.
The function $\mathbf{f}: \mathcal{S} \rightarrow \mathcal{G}$ is a black-box simulator mapping state to achieved goal.
A deterministic transition function $\mathbf{T}: \mathcal{S} \times \mathcal{A} \rightarrow \mathcal{S}$ governs state transitions, with $\mathbf{s}_0$ representing the initial state.
The reward function $r: \mathcal{G} \times \mathcal{G} \rightarrow \mathbb{R}$ measures the discrepancy between the achieved goal and the episode’s target goal.
Finally, $H$ denotes the maximum episode length, after which the state is reset, or the state is reset earlier upon achieving the target goal.

A state at step count $t \in \{0,1,\dots,H-1\}$ within our framework is defined as $\mathbf{s}_t \coloneqq [W_{1,t}, W_{2,t},\dots, C_{\text{load},t}]$, explicitly representing $L$ circuit parameters given the fixed circuit topology.
Here, $W$ denotes the transistor width, and $C$ denotes the capacitance.
Unlike previous GCRL approaches~\cite{autockt, cao2024roseoptrobustefficientanalog} that represent actions via discrete parameter adjustments, such as modifying transistor multipliers or discretizing capacitances, we define the action $\mathbf{a}_t \coloneqq \mathbf{s}_{t+1}-\mathbf{s}_t$, capturing continuous adjustments of the circuit parameters at each step.
The deterministic transition function is straightforwardly defined as $\mathbf{T}(\mathbf{s}_t, \mathbf{a}_t) \coloneqq \mathbf{s}_t + \mathbf{a}_t$.
This continuous formulation enables finer control and enhances the accuracy and adaptability in exploring the design space.

The episode target goal $\mathbf{g}$ is defined as the concatenation ($\|$) of two vectors:
\begin{equation}
\label{target_goal}
\begin{aligned}
\mathbf{g} &\coloneqq \hat{\mathbf{z}} \| [1,1], \quad \text{where } \hat{\mathbf{z}} \sim p_{\text{goal}}(\hat{\mathbf{z}}).
\end{aligned}
\end{equation}
Here, $p_{\text{goal}}$ denotes the distribution over $M$ target specifications for the circuit (e.g., bandwidth, gain). The appended binary vector $[1, 1]$ represents desired satisfaction indicators, one for the nominal corner and one for all corners.
While $\mathbf{g}$ remains constant throughout an episode, the achieved goal $\mathbf{g}_t$ at each step is computed through circuit simulation $\mathbf{f}$:
\begin{equation}
\label{achieved_goal}
\begin{aligned}
\mathbf{g}_t(\mathbf{s}_t,\mathbf{a}_t) &= \mathbf{f}(\mathbf{s}_t + \mathbf{a}_t) = \mathbf{z}_t \,\|\, [D^{\text{nom}}_t, D_t] \\
&\coloneqq
\begin{cases}
\mathbf{z}^0_t \,\|\, [0, 0], & \text{if } s = 1, \\
\text{Worst}(Z_t) \,\|\, [1, D_t], & \text{if } s \ne 1.
\end{cases}
\end{aligned}
\end{equation}
The binary vector $[D^{\text{nom}}_t, D_t] \in \{0,1\}^2$ encodes whether the metric vector $\mathbf{z}_t$ satisfies the target specifications in the nominal corner ($D^{\text{nom}}_t = 1$) and in all corners ($D_t = 1$), respectively.
$\mathbf{z}^0_t \in \mathbb{R}^M$ denotes the vector of performance metrics measured at the nominal corner, used when full-corner simulation is skipped.
Otherwise, when computing $\mathbf{g}_t$, we form the matrix
$Z_t = [\mathbf{z}_t^{1}, \mathbf{z}_t^{2}, \ldots, \mathbf{z}_t^{N}]^\top \in \mathbb{R}^{N \times M}$, which contains $N$ metric vectors. 
Each $\mathbf{z}^k_t \in \mathbb{R}^M$ records the performance metrics at the $k^\text{th}$ corner at time step $t$, and $N$ denotes the total number of PVT corners excluding the nominal one used during training.
The function $\text{Worst}(\cdot)$ computes a column-wise worst-case (e.g., max or min depending on the metric) across the $N$ corners, returning a single $M$-dimensional vector.

We specifically utilize the worst-case metrics across all corners  to formulate $\mathbf{g}_t$, rather than employing all intermediate metrics.
Instead, to enhance decision-making clarity, we introduce a binary indicator $D^{\text{nom}}_t$ within the goal formulation.
This indicator explicitly signals whether the agent is operating on the nominal corner trajectory ($s\!=\!1$) or has transitioned to considering all corners ($s\neq\!1$).
Incorporating this explicit trajectory information is essential, as the agent would otherwise be entirely invariant to PVT variations.
Note that the achieved goal $\mathbf{g}_t$ is used solely to compute the reward and does not directly serve as input to any neural network unless it undergoes goal relabeling. 
Additional details regarding the goal relabeling process are discussed in Section~\ref{CHER}.

While previous GCRL-based approaches~\cite{autockt, cao2024roseoptrobustefficientanalog} predominantly employ PPO~\cite{ppo}, we adopt Soft Actor-Critic (SAC)~\cite{sac} as our reinforcement learning optimizer.
The choice of SAC is primarily motivated by the high computational overhead associated with multi-corner simulations.
As discussed in Section~\ref{section: related works}, PPO, being an on-policy algorithm, cannot effectively benefit from parallel rollout without specialized tools.
Thus, we select SAC, an off-policy algorithm known for its superior robustness and stability in high-dimensional action spaces, especially compared to methods such as Deep Deterministic Policy Gradient (DDPG)~\cite{sac, ddpg}.
Within our framework, SAC trains a stochastic goal-conditioned actor $\boldsymbol{\pi}_\theta(\mathbf{a}_t|\mathbf{s}_t,\mathbf{g})$ and two critic functions $Q_{\phi_1}(\mathbf{s}_t, \mathbf{a}_t, \mathbf{g})$ and $Q_{\phi_2}(\mathbf{s}_t, \mathbf{a}_t, \mathbf{g})$.

\subsection{Pareto Dominance Goal Sampling}
\label{pgds}
\begin{algorithm}[t]
\caption{Pareto Dominance Goal Sampling Strategy}
\label{alg:goal sampling method}
\begin{algorithmic}[1]
\STATE \textbf{Input:} Actor $\boldsymbol{\pi}_\theta$, Critics $Q_{\phi_1}$, $Q_{\phi_2}$,
achieved goal buffer $\mathcal{R}$, reset stage $\mathbf{s}_0$
\STATE \textbf{Output:} Sampled goal $\mathbf{g}$
\IF{$|\mathcal{R}|\leq N_\text{uniform}$}
    \STATE Sample $\mathbf{g} \sim \text{Uniform}(\mathcal{G})$
    \RETURN $\mathbf{g}$
\ENDIF
\FOR{$i=1,2,\dots N_g$}
    \REPEAT
        \STATE Sample $\mathbf{g}_i\sim \text{Uniform}(\mathcal{G})$
    \UNTIL $\mathbf{g}_i$ is not \textsc{ParetoDominated} by $\mathcal{R}$
    \STATE Sample $\mathbf{a}_i \sim \boldsymbol{\pi}_\theta(\cdot|\mathbf{s}_0,\mathbf{g}_i)$
    \STATE $Q_i \gets \frac{1}{2}\left[Q_{\phi_1}(\mathbf{s}_0,\mathbf{a}_i,\mathbf{g}_i)+ Q_{\phi_2}(\mathbf{s}_0,\mathbf{a}_i,\mathbf{g}_i)\right]$
\ENDFOR
\STATE $\mathbf{p} \gets \text{Softmax}(-[Q_1, Q_2, \dots,Q_{N_g}]/T)$
\STATE Sample $k \sim \text{Categorical}(\mathbf{p})$
\RETURN $\mathbf{g}_k$
\end{algorithmic}
\end{algorithm}

A common brute-force approach for sampling target goal is to uniformly draw goal within predefined target ranges for each metric~\cite{settaluri2020autocktdeepreinforcementlearning, cao2024roseoptrobustefficientanalog}.
However, this strategy often proves inefficient for learning: in early training stages, most sampled goals are overly ambitious, while in later stages, they tend to be too trivial~\cite{andrychowicz2018hindsightexperiencereplay,automatic_curriculum}.
Curriculum learning methods~\cite{automatic_curriculum} address this by gradually increasing the difficulty of sampled goals, aiming to provide training tasks that are neither too easy nor too difficult—thereby maximizing learning efficiency, especially in sparse reward environments.
However, when employing dense rewards, assigning difficult goals does not negatively impact learning as severely as in sparse reward scenarios since challenging goals still provide sufficient training signals through dense feedback.

Alternatively, we propose Pareto Dominance Goal Sampling (PDGS), an adaptive goal selection strategy that leverages both historical goal-achievement data and the current policy to construct a dynamic curriculum.
The process begins by identifying $N_g$ candidate goals that exhibit moderate to high difficulty—specifically, those that are not \textit{Pareto-dominated} by previously achieved specifications.
A candidate goal is considered \textit{Pareto-dominated} if there exists a previously satisfied goal whose specification values are all superior to those of the candidate.
From the set of $N_g$ non-\textit{Pareto-dominated} candidate goals, PDGS selects a single goal.
To avoid sampling overly easy targets—even within this filtered set—it prefers goals that are more difficult.
Goal difficulty is quantified using the mean value estimated by two parametric Q-functions, $Q_{\phi_1}$ and $Q_{\phi_2}$.
Lower Q-values indicate greater difficulty, implying that the selected goal resides near or beyond the current policy's knowledge boundary.


However, greedily sampling goals based on Q-estimates can be problematic when the estimates are imprecise or the goals are overly difficult.
A single hardest goal may stall learning if it is out of reach or incorrectly estimated.
Instead, PDGS adopts a soft, non-greedy sampling strategy by drawing from a distribution that assigns higher probabilities to goals with lower Q-estimates.
While various methods exist for introducing stochasticity into the sampling process, we adopt an approach that treats the negative Q-estimates as logits and computes categorical probabilities via softmax, based on superior empirical performance.
In this way, it favors difficult goals without always locking onto one that is potentially unsolvable or misleading under the current Q-function approximation. This procedure is analogous to self-normalized importance sampling, where each candidate is assigned an importance weight and sampling occurs in proportion to these normalized weights.


Algorithm~\ref{alg:goal sampling method} details the goal sampling procedure. The temperature parameter $T$ controls the smoothness of the softmax distribution used for non-greedy selection. As $T\! \rightarrow \!\infty$, the distribution approaches uniform sampling; as $T \!\rightarrow\! 0$, it converges to greedy selection.
PGDS is activated only after the number of achieved goals exceeds $N_\text{uniform}$ to promote exploration during early training.
After this phase, goals are sampled from the Pareto frontier—an effective region for policy training, as these goals are nontrivial.
By focusing on such goals, PDGS establishes an automatic learning curriculum that enhances sample efficiency.  A concise visual representation of the PGDS method is provided in step 1 of Fig.~\ref{fig:overview}.
\subsection{PVT-aware Hierarchical Reward with PVT-consistency}
\label{reward}
\begin{algorithm}[t]
    \caption{Overall Training Process}
    \label{arg: training pseudocode}
    \begin{algorithmic}[1]
        \STATE \textbf{Input:} Initial state $\mathbf{s}_0$
        \STATE \textbf{Initialize:} Actor parameters $\theta$, critic parameters $\phi_1$,$\phi_2$, target critic parameters $\phi'_2$,$\phi'_2$, replay buffer $\mathcal{D}$, episode buffer $\mathcal{E}$, and achieved goal buffer $\mathcal{R}$
        \FOR{each episode}
            \STATE $\mathbf{g} \gets \textsc{PGDS}(\boldsymbol{\pi}_\theta$, $Q_{\phi_1}$, $Q_{\phi_2}$, $\mathcal{R}$, $\mathbf{s}_0$)
            \STATE $\mathcal{E} \gets \emptyset$
            \FOR{t = $0,1,,\dots,H-1$} 
                \STATE Sample $\mathbf{a_t} \sim \pi_\theta\left(\cdot | \mathbf{s_t}, \mathbf{g}\right)$ 
                \STATE $(\mathbf{s}_{t+1},r_t,\mathbf{g}_t) \gets \textsc{SoF\_EnvStep}(\mathbf{s}_t,\mathbf{a}_t)$
                \STATE Store $(\mathbf{s}_t, \mathbf{a}_t, r_t, \mathbf{s}_{t+1}, \mathbf{g}_t, \mathbf{g},t )$ in $\mathcal{E}$ 
                \IF{$D_t$}
                    \STATE Store $\mathbf{g}$ in $\mathcal{R}$ and Break (Goal achieved)
                \ENDIF
            \ENDFOR
            \FOR{each transition $(\mathbf{s}_t, \mathbf{a}_t, r_t, \mathbf{s}_{t+1}, \mathbf{g}_t, \mathbf{g},t)$ in $\mathcal{E}$}
                \STATE $t'\sim \text{Uniform}([t+1:|\mathcal{E}|])$
                \STATE $r'_t = r_{R'}\left(\mathbf{g}_t, Z_t, \mathbf{g}_{t'}\right)$ 
                \STATE Store $(\mathbf{s}_t, \mathbf{a}_t, r'_t, \mathbf{s}_{t+1}, \mathbf{g}_{t'})$, $(\mathbf{s}_t, \mathbf{a}_t, r_t, \mathbf{s}_{t+1}, \mathbf{g})$ in $\mathcal{D}$
            \ENDFOR
            
            \FOR{each gradient step}
                \STATE Sample $(\mathbf{s_t}, \mathbf{a_t}, r_t, \mathbf{s_{t+1}}, \mathbf{g})$ $\sim$ $\mathcal{D}$ (batch size = $\mathcal{B}$) 
                \STATE Update actor $\pi_\theta$, critics $Q_{\phi_1}$, $Q_{\phi_2}$,  and target critics $Q_{\phi'_1}$, $Q_{\phi'_2}$ with batched transitions
            \ENDFOR
        \ENDFOR
    \end{algorithmic}
\end{algorithm}
Following the principles of the SoF approach, it is necessary to design a stage-aware reward function. A key constraint is that the reward obtained from the first stage—where simulations are run only at the nominal corner—must not exceed the reward computed in the second stage, which evaluates across all corners. Assuming that the target goal constraints are lower bounded, the reward is computed from the tuple $(\mathbf{g}_t, Z_t, \mathbf{g})$ as follows:
\begin{equation}
\begin{aligned}
&r_R(\mathbf{g}_t, Z_t, \mathbf{g}) =\\
&\begin{cases}
R \left(1 - \psi(\mathbf{z}_t, \hat{\mathbf{z}})\right) + R_{\min} \psi(\mathbf{z}_t, \hat{\mathbf{z}}) - \alpha, & \text{if } s = 1, \\
R \, \psi(\mathbf{z}_t, \hat{\mathbf{z}}) - \alpha \, \sigma(Z_t), & \text{if } s = 2, \\
R_{\max} - \alpha \, \sigma(Z_t), & \text{if } s = 3.
\end{cases}
\end{aligned}
\end{equation}
It is worth mentioning that the stage can be identified by $[D^{\text{nom}}_t, D_t]$ contained in $\mathbf{g}_t$. $R_{\max} \!\geq\! 0$ denotes the maximum achievable reward, assigned when the agent satisfies all specifications.
The values $R \!\leq\! 0$ and $R_{\min} \!\leq\! R$ serve as interpolation anchors in the second and first stages, respectively.
An additional penalty term $\sigma(Z_t)$, scaled by a non-negative weight $\alpha \!\geq\! 0$, is introduced to promote consistent performance under PVT variations.
The function $\psi: \mathbb{R}^M \!\times\! \mathbb{R}^M \rightarrow [0, 1]$ is an aggregation function that quantifies the normalized difference between the observed output $\mathbf{z}_t$ and the target specification $\mathbf{z}$.
It satisfies the boundary conditions $\psi(\hat{\mathbf{z}}, \hat{\mathbf{z}}) = 0$ and $\psi(\mathbf{0}, \hat{\mathbf{z}}) = 1$, where $\mathbf{0} \in \mathbb{R}^M$ denotes an all-zero vector.
The metric is defined as:
\begin{equation}
    \begin{aligned}
    \label{normalize_function}
    \psi(\mathbf{z}_t,\hat{\mathbf{z}}) &\coloneqq
    M^{-1} \mathbf{h}(\mathbf{z}_t, \hat{\mathbf{z}})^\top \mathbf{1},\\
    \text{where} \quad \mathbf{h}(\mathbf{z}_t, \hat{\mathbf{z}}) &\coloneqq 
    \frac{\tanh\left(\eta \left(\mathbf{1} - \mathbf{z}_t \oslash \hat{\mathbf{z}} \right)\right)}{\tanh(\eta)}.
    \end{aligned}
\end{equation}
$\eta$ is a scalar hyperparameter that prevents numerical overflow and controls the sensitivity of the monotonically decreasing normalizer $\mathbf{h}$.
The operator $\oslash$ denotes element-wise division, and $\mathbf{1} \in \mathbb{R}^M$ is an all-one vector.

Consequently, the reward in the first stage interpolates between $R_{\min}$ and $R$ until the agent meets the target specifications.
In this context, $R$ represents the intermediate maximum reward assigned for successfully achieving the target goal in the first stage.
On the other hand, the second stage reward interpolates between $R$ and $0$ with interpolation coefficient $\psi(\mathbf{z}_{t},\hat{\mathbf{z}})$.
The interpolation ensures that the reward calculated from the full corner simulations is greater than that calculated only from the nominal corner, successfully satisfying the aforementioned constraint, thereby urging the agent not to be complacent in the nominal corner.
Note that this formulation assumes the target goal constraints are lower bounded (e.g., Gain, UGBW).
For metrics that have an upper bound (e.g., Power, Delay), we reverse the sign without loss of generality.


In addition to the principled reward formulation, we introduce a penalty term $\sigma(Z_t)$ as a sub-objective to promote consistency across PVT corners:
\begin{equation}
    \label{pvt consistency term}
    \sigma(Z_t)=\frac{1}{MN}\sum^N_{i=1}\sum^M_{j=1}\left(\frac{z^i_{t}[j]}{z^{0}_{t}[j]}-1\right)^2.
\end{equation}
We not only encourage the agent to meet the target specifications across all corners, but also explicitly minimize deviation from the nominal corner performance. This dual objective reflects a practical design goal—ensuring robust performance under PVT variations.
\subsection{Conservative Hindsight Experience Replay}
\label{CHER}
To remedy the lowered sample quality caused by the SoF simulation framework, we leverage the strength of HER, but with modifications in computing the virtual reward for relabeled goals.
HER is particularly effective in sparse reward settings, and our hierarchical reward formulation—defined using three stages over distinct ranges—naturally introduces structured sparsity into the learning signal.
This hierarchy induces a reward landscape with well-separated levels of difficulty, resembling a sparse reward structure where positive feedback is only given upon reaching progressively stricter target conditions.
As a result, relabeled goals that fall within different stages provide more informative and interpretable feedback.
This structure inherently introduces discreteness into the dense reward signal, as it assigns qualitatively different feedback across separate regions of the state and goal space.

Formally, we add a synthetic transition $(\mathbf{s}_t, \mathbf{a}_t, r'_t, \mathbf{s}_{t+1}, \mathbf{g}_{t'})$ per single original transition $(\mathbf{s}_t, \mathbf{a}_t, r_t, \mathbf{s}_{t+1}, \mathbf{g})$ into the replay buffer by relabeling the target goal $\mathbf{g}$ with an achieved goal of an arbitrary future state $\mathbf{g}_{t'}$ in the episode.
The virtual reward $r'_t$ is computed conservatively as
$r'_t = r_{R'}(\mathbf{g}_t, Z_t, \mathbf{g}_{t'})$,
where $R'$ satisfies the inequality $R' \leq R \leq 0$.
This change of variable enforces the inequality
\begin{equation}
r_{R'}(\mathbf{g}_t, Z_t, \mathbf{g}_{t'}) \leq r_{R}(\mathbf{g}_t, Z_t, \mathbf{g}_{t'}),
\end{equation}
thereby providing a conservative learning signal for the relabeled transitions.
This conservative setting encourages the agent to adopt safer policies~\cite{pessimistic_reward}, improving its robustness and performance across significantly varying PVT corner conditions.
In practice, we implement this using an episode buffer $\mathcal{E}$, from which both original and relabeled transitions are stored in the replay buffer for off-policy learning.
Note that our Conservative HER (CHER) no longer requires additional simulations or surrogate model~\cite{cronus} to synthesize virtual transition.
The complete training procedure is detailed in Algorithm~\ref{arg: training pseudocode}.
\subsection{Deployment Strategy}
In the deployment stage, the agent is no longer trained and deterministically selects actions given previously unseen target goals.
This highlights the GCRL agent's ability to generalize to new target goals that were not encountered during training.
Note that while the agent samples actions from the probabilistic model during training, it selects the mean action from the Gaussian distribution during deployment, resulting in deterministic behavior.
\section{Experimental Results}
\label{experiments}
\subsection{Experimental Setups}




\begin{table*}[t]
  \centering
  \caption{Circuit Design Parameters}
  \label{tab:design_parameters}
  \resizebox{\linewidth}{!}{%
    \begin{tabular}{c c c c c c c c c c c}
      \toprule
      \multirow{2}{*}{\textbf{Circuit}}
        & \multirow{2}{*}{\textbf{\#Params}}
        & \multicolumn{3}{c}{\textbf{PMOS}}
        & \multicolumn{3}{c}{\textbf{NMOS}}
        & \multirow{2}{*}{{V$_b$} (V)}
        & \multirow{2}{*}{{R} (k$\Omega$)}
        & \multirow{2}{*}{{C} (pF)} \\
      \cmidrule(lr){3-5} \cmidrule(lr){6-8}
      & 
        & {W} ($\mu$m) & {L} ($\mu$m) & {M}
        & {W} ($\mu$m) & {L} ($\mu$m) & {M}
        &  &  &  \\
      \midrule
      TSA   & 7
        & [0.5,\,100]$^2$ & 0.3$^2$ & 1$^2$
        & [0.5,\,100]$^4$ & 0.3$^4$ & 1$^4$
        & —        & —         & [0.1,\,10]  \\
      CMA   & 10
        & [0.5,\,100]$^5$ & 0.3$^5$ & 1$^5$
        & [0.5,\,100]$^3$ & 0.3$^3$ & 1$^3$
        & —        & —         & [0.1,\,10]$^2$  \\
      Comp  & 6
        & [0.5,\,100]$^3$ & 0.6$^3$ & 1$^3$
        & [0.5,\,100]$^3$ & 0.3$^3$ & 1$^3$
        & —        & —         & —         \\
      LDO   & 17
        & [1,\,100]$^2\times$[10,\,100] 
        & [0.5,\,2]$^2\times$[0.5,\,1] & 1$^2\times$[100,\,2000]
        & [1,\,100]$^3$ & [0.5,\,2]$^3$ & 1$^3$
        & [0.9,\,1.4]
        & [0.4,\,10]
        & [0.2,\,10]$\times$[20,\,550]  \\
      \bottomrule
    \end{tabular}%
  }
    \vspace{-5pt}
\end{table*}

\begin{figure*}
  \centering
  %
  \begin{subfigure}[t]{0.22\textwidth}
    \centering
    \includegraphics[width=\linewidth]{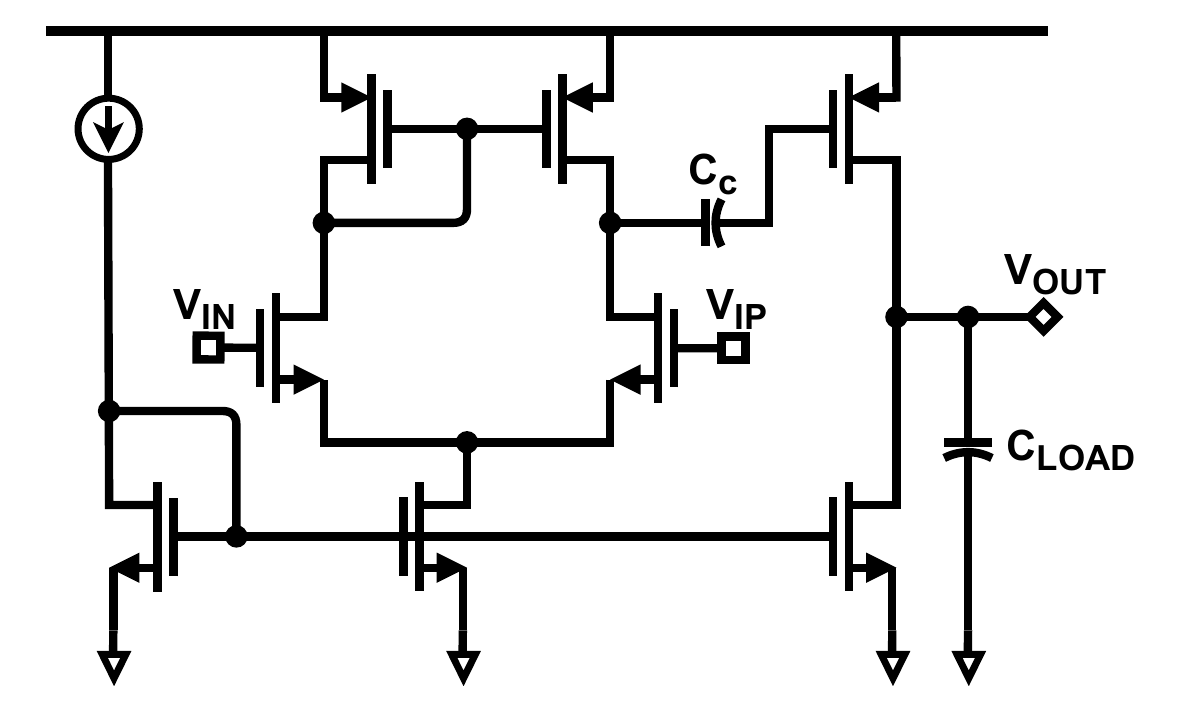}
    \caption{Two‑Stage Op.\ Amp}
    \label{fig:tso}
  \end{subfigure}\hfill
  \begin{subfigure}[t]{0.36\textwidth}
    \centering
    \includegraphics[width=\linewidth]{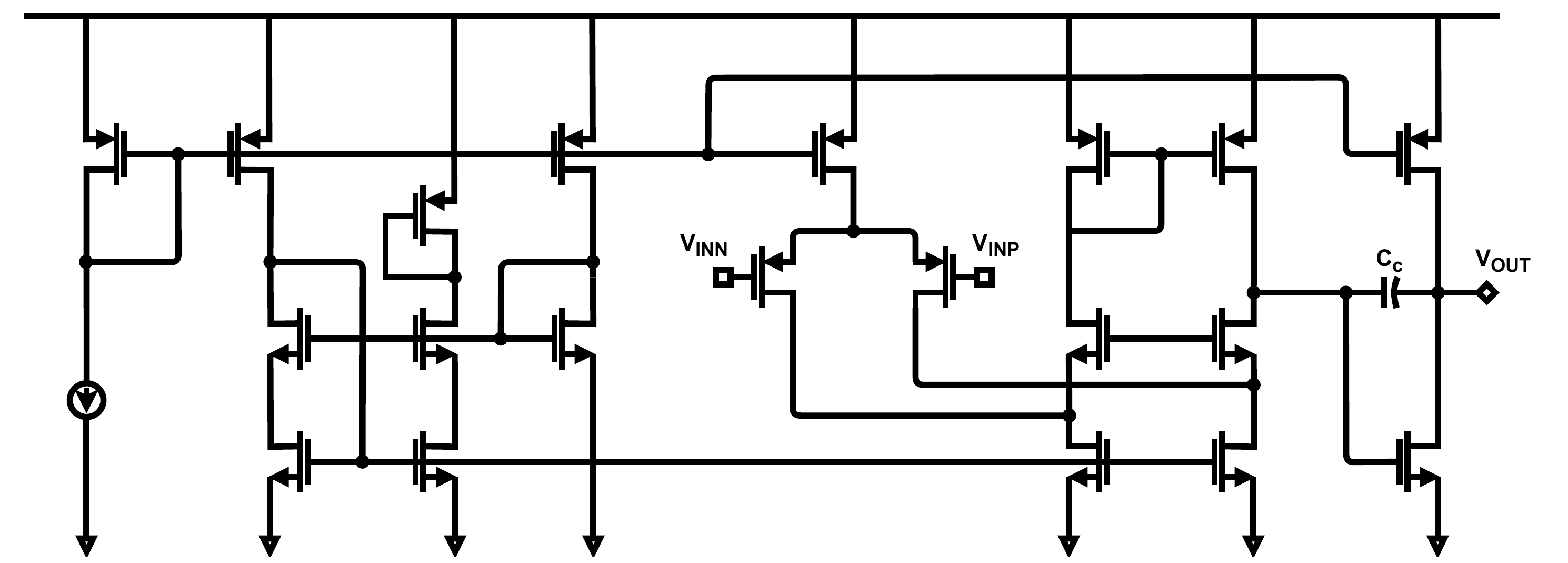}
    \caption{Cascode Miller‑Comp.\ Amp}
    \label{fig:cma}
  \end{subfigure}\hfill
  \begin{subfigure}[t]{0.22\textwidth}
    \centering
    \includegraphics[width=\linewidth]{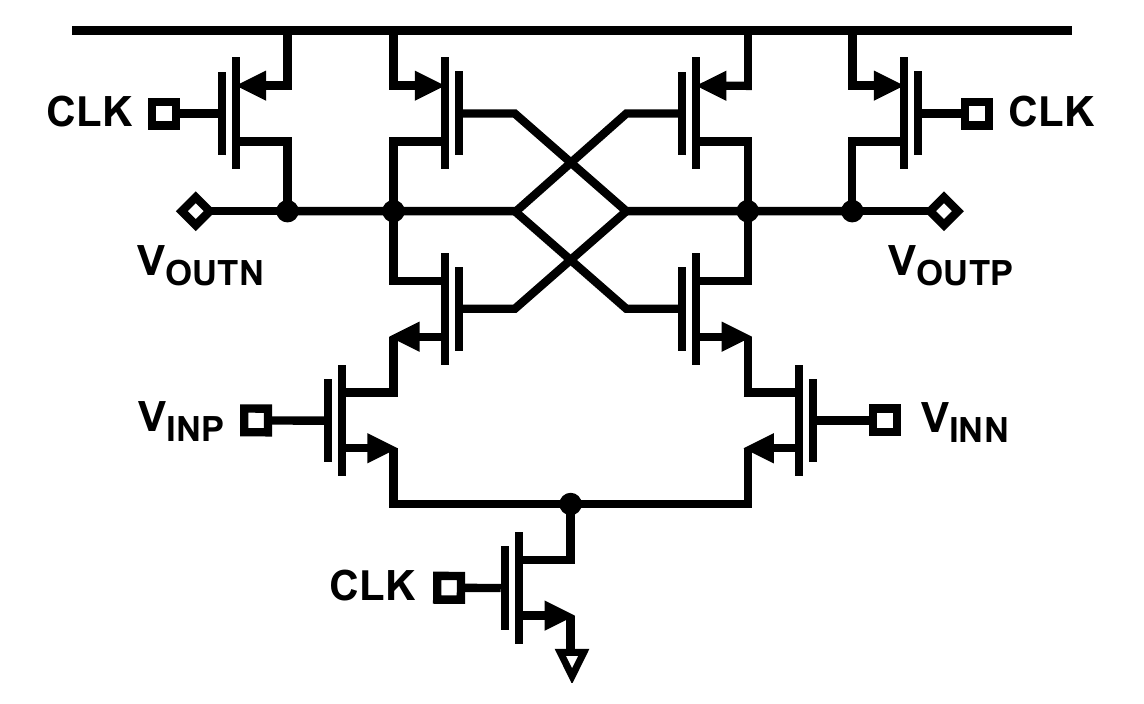}
    \caption{Comparator}
    \label{fig:comp}
  \end{subfigure}\hfill
  \begin{subfigure}[t]{0.18\textwidth}
    \centering
    \includegraphics[width=\linewidth]{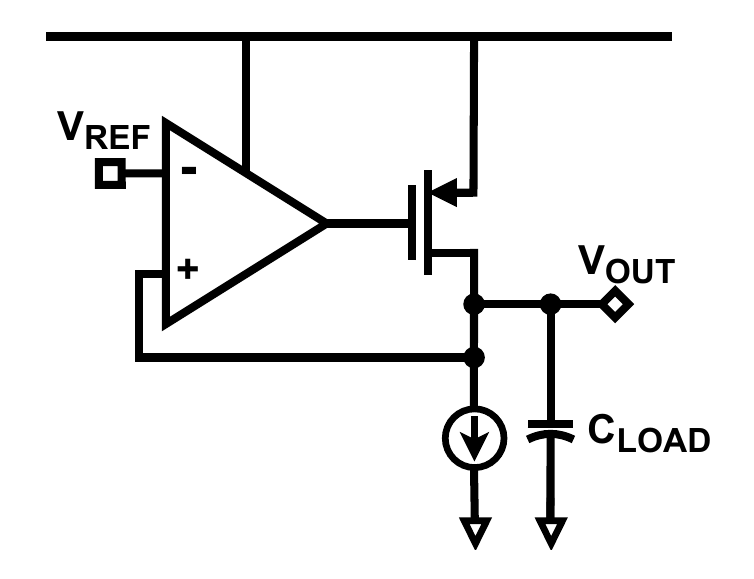}
    \caption{LDO}
    \label{fig:ldo}
  \end{subfigure}

  \caption{Topologies of the four benchmark circuits.}
  \label{fig:circuits}
  \vspace{-10pt}
\end{figure*}

\begin{table}[t]
  \centering
  \small
  \caption{Circuit Target Specifications}
  \label{tab:target_specs}
    \begin{tabular}{c l c l}
      \toprule
      \textbf{Circuit} & \textbf{Metric}                     & \textbf{Bound}   & \textbf{Target Range}      \\
      \midrule
      \multirow{6}{*}{TSA}
        & Gain (dB)                           & $\geq$     & [46,\,52]           \\
        & PM ($^\circ$)                       & $\geq$     & 60                  \\
        & UGBW (MHz)                          & $\geq$     & [1,\,20]            \\
        & $V_{\text{swing}}$ (V)              & $\geq$     & [0.2,\,0.3]         \\
        & Power (mW)                          & $\leq$   & [3.3,\,33]          \\
        & $T_{\text{settle}}$ ($\mu$s)       & $\leq$   & [2.0,\,10.0]        \\
      \midrule
      \multirow{6}{*}{CMA}
        & Gain (dB)                           & $\geq$     & [73,\,76]           \\
        & PM ($^\circ$)                       & $\geq$     & 85                  \\
        & UGBW (MHz)                          & $\geq$     & [4,\,6]             \\
        & $V_{\text{swing}}$ (V)              & $\geq$     & [1.4,\,1.6]         \\
        & Power (mW)                          & $\leq$   & [9.9,\,33]          \\
        & $T_{\text{settle}}$ ($\mu$s)       & $\leq$   & [0.15,\,0.3]        \\
      \midrule
      \multirow{8}{*}{LDO}
        & $V_{\text{drop}}$ (mV)             & $\leq$   & [250,\,350]         \\
        & PM$^\text{min}$/PM$^\text{max}$ ($^\circ$)     & $\geq$     & 60                  \\
        & PSRR$_{\leq10kHz}^\text{min}$ (dB)             & $\leq$   & [-20,\,–10]         \\
        & PSRR$_{\leq1MHz}^\text{min}$ (dB)              & $\leq$   & [-15,\,–10]         \\
        & PSRR$_{>1MHz}^\text{min}$ (dB)             & $\leq$   & [-20,\,–10]         \\
        & PSRR$_{\leq10kHz}^\text{max}$ (dB)             & $\leq$   & [-25,\,–20]         \\
        & PSRR$_{\leq1MHz}^\text{max}$ (dB)              & $\leq$   & [-15,\,–10]         \\
        & PSRR$_{>1MHz}^\text{max}$ (dB)             & $\leq$   & 0                   \\
      \midrule
      \multirow{2}{*}{Comp}
        & Delay (ns)                          & $\leq$   & [0.1,\,0.2]         \\
        & Switching Power (nW)               & $\leq$   & [0.05,\,0.15]       \\
      \bottomrule
    \end{tabular}
    \vspace{-10pt}
\end{table}
We demonstrate the effectiveness of our framework with four benchmark circuits in Fig.~\ref{fig:circuits}: a two-stage operational amplifier (TSA), a cascode miller-compensated amplifier with bias circuit (CMA) ~\cite{analoggym,fmc_3s_amp}, a comparator (COMP), and a low-dropout regulator (LDO) \cite{sky130_ldo_rl}.
Each benchmark is trained for $12K$, $24K$, $8K$, and $12K$ steps, respectively.
The circuits are implemented using open-source PDKs, where LDO is based on the SKY130 PDK and the others are implemented with the GF180MCU PDK\cite{gf180mcu_pdk}.
The target specifications listed in Table~\ref{tab:target_specs} define the desired performance ranges. 
The symbol $\geq$ indicates that a specification must exceed the target goal to be considered successful, while $\leq$ means the specification must remain below the target threshold to succeed.
The corresponding design parameters optimized during training are summarized in Table~\ref{tab:design_parameters}.

For the TSA, CMA, and COMP benchmarks, we conducted training across 17 PVT corners, including one nominal corner, [TT, 3.3V, 27$^\circ$C], and 16 extreme corners \{FF, SS, SF, FS\} $\times$ \{3.0V, 3.6V\} $\times$ \{-40$^\circ$C, 125$^\circ$C\}.
Extreme corner cases have $\pm$0.3V deviation from the nominal supply voltage 3.3V. 
Subsequently, deployment is conducted across a full grid of 45 PVT corners, spanning \{TT, FF, SS, SF, FS\} $\times$ \{3.0V, 3.3V, 3.6V\} $\times$ \{-40$^\circ$C, 27$^\circ$C, 125$^\circ$C\}.
For the LDO benchmark, both training and evaluation are performed over 9 PVT corners, including one nominal corner,[TT, 2.0V, 27$^\circ$C], and 8 extreme corners \{FF, SS, SF, FS\} $\times$ \{-40$^\circ$C, 125$^\circ$C\}, without supply voltage variation.
Following the configuration in~\cite{sky130_ldo_rl}, the PSRR and PM are evaluated under two load conditions: $I_L^{\min} = 10\mu\text{A}$ and $I_L^{\max} = 10\text{mA}$, as performance depends on load current.

To facilitate reproducibility, all experiments are conducted using the open-source Ngspice circuit simulator and open-source PDKs. The actor and critic networks in our GCRL framework are modeled as fully connected neural networks, with hidden dimensions of $[256, 256, 256, 256]$ for the actor and $[256, 256, 128]$ for the critics. Both networks employ the $\tanh$ activation function.

Prior to training, we initialize the environment with a fixed state $\mathbf{s}_0$, which serves as the reset state for each episode. This state is selected by evaluating 50 random candidates in the nominal corner and choosing the one that yields the highest reward, independent of the goal. An exception is made for the LDO benchmark, where $\mathbf{s}_0$ is initialized to the optimized values reported in~\cite{sky130_ldo_rl}.

We configure our experiments with the following hyper-parameters. For the RL algorithm, we set learning rate $lr = 0.003$, batch size $\mathcal{B}=256$, discount factor $\gamma = 0.8$, goal sampling temperature $T = 5.0$, number of goal candidates $N_g = 16$, and uniform goal sampling threshold $N_{\text{uniform}} = 4$.
The episode horizon is fixed to $H = 30$ during both training and deployment. We set the normalizer scale to $\eta = 0.1$, the maximum reward to $R_{\max} = 30.0$, the intermediate reward to $R = -1$, the conservative counterpart to $R' = -3$, and the minimum reward to $R_{\min} = -6$.
Experiments are conducted on a 64-core AMD CPU workstation.

\subsection{Metrics in Experiments}
\begin{table*}[!t]
\caption{Performance Comparisons}
\label{tab:performance_metrics}
\resizebox{\textwidth}{!}{%
\begin{tabular}{c|c|c|ccc|ccc|ccc|ccc}
\toprule
\multirow{2}{*}{Method}
  & \multirow{2}{*}{PVT?}
  & \multirow{2}{*}{Generalize?}
  & \multicolumn{3}{c|}{TSA}
  & \multicolumn{3}{c|}{CMA}
  & \multicolumn{3}{c|}{LDO}
  & \multicolumn{3}{c}{COMP} \\
\cmidrule(lr){4-6} \cmidrule(lr){7-9} \cmidrule(lr){10-12} \cmidrule(lr){13-15}
 &  & 
 & SR(\%)      & S$_\text{sim}$    & S$_\text{dev}$(m)
 & SR(\%)      & S$_\text{sim}$    & S$_\text{dev}$(m)
 & SR(\%)      & S$_\text{sim}$    & S$_\text{dev}$(m)
 & SR(\%)      & S$_\text{sim}$    & S$_\text{dev}$(m) \\
\midrule
RoSE-Opt~\cite{cao2024roseoptrobustefficientanalog}           & Full    & Yes & 70.5  &  3.6  &   181 & 44.8  &  1.1 &  202 & 42.0  & 2.0  &   173 & 48.0 &  3.75 & 206  \\
RobustAnalog~\cite{shi2022robustanalogfastvariationawareanalog}       & Partial & No  &  0.0  &  0.0  &  N/A & 39.3  &  5.4 &  195 &  0.0  & 0.0  &  N/A & \textbf{78.3}$^\text{a}$ & 12.2 & 210  \\
AutoCkt~\cite{settaluri2020autocktdeepreinforcementlearning}           & No      & Yes &  3.2  &  2.6  &  242 &  0.0  &  0.0 &  N/A &  0.0  & 0.0  &  N/A &  5.2 &  6.5  & 223  \\
BO~\cite{bo_analog}                & No      & No  &  0.0  &  0.0  &  N/A &  0.0  &  0.0 &  N/A &  0.0  & 0.0  &  N/A &  0   &  0    &  N/A \\
RL Baseline        & Full    & Yes & 77.6  & 18.5  &  184 & 35.0  &  3.8 &  203 & 87.3  & 9.1  &  171 & 69.0 & 22.7  & 200  \\
\rowcolor{gray!15}
PPAAS               & Full    & Yes & \textbf{92.6} & \textbf{20.6} & 191 & \textbf{89.3} & \textbf{9.0}  & 192 &   88.4   &  \textbf{12.6}   &  175  & 69.6 & \textbf{33.9} & 204  \\
\rowcolor{gray!15}
PPAAS ($\alpha=10$)  & Full    & Yes & 87.5  & 17.8  & \textbf{175} &   78.2 &  7.7   &  \textbf{189} &   \textbf{91.6}   &  10.5   &  \textbf{160}  & 73.5 & 26.6  & \textbf{198} \\
\hline
Improvement  & --    & -- & 1.3$\times$  & 5.7$\times$  & +3\% &   2.0$\times$ &  1.8$\times$   &  3\% &   2.2$\times$   &  6.3$\times$   &  +8\%  & 0.9$\times$ & 2.8$\times$  & +4\% \\
\bottomrule
\end{tabular}%
}
\caption*{
\footnotesize $^\text{a}$ Though two specifications suggest two corners, goal-dependent variation leads us to select 8 from the full 16. Otherwise, the success rate is 0.\\
}
\vspace{-20pt}
\end{table*}

We evaluate the performance of our framework using a primary metric, the success rate (SR), and two sub-metrics: the simulation efficiency score (S$_\text{sim}$) and the normalized deviation score (S$_\text{dev}$).

The success rate is defined as $\text{SR} \!=\! \frac{N_{\text{success}}}{N_{\text{eval}}}$, where $N_{\text{success}}$ denotes the number of successful episodes out of $N_{\text{eval}} \!=\! 500$ evaluation episodes. The evaluation target goals are uniformly pre-sampled within the ranges specified in Table~\ref{tab:target_specs} and are fixed for each benchmark, whereas target goals during training are sampled adaptively using PGDS. An episode is considered successful if the agent reaches the target goal before reaching the maximum horizon $H$.
Since all models are trained with the same number of environment steps for each benchmark, the success rate provides a direct measure of sample efficiency—i.e., how effectively each method utilizes its training budget to solve given target goals. Note that the success rate serves as a proxy for generalization ability as well.

The simulation efficiency score, S$_\text{sim}$, accounts for the total number of simulations required to train the agent, thereby reflecting both the simulation cost and the generalization ability of the agent.
It is defined as
S$_\text{sim} = \frac{\text{SR}}{\#\text{sim}}\times10^6$,
where $\#\text{sim}$ denotes the total number of simulations performed during training.
Note that a single training step may involve either one simulation under the nominal corner or $N$ simulations across all PVT corners.
The normalized deviation score S$_{\text{dev}}$ is computed as the mean of the penalty term $\sigma(Z_t)$ at the final transition of each successful episode.
This term quantifies the deviation of the achieved specifications from the nominal corner, thereby indicating the degree of consistency across PVT corners.
All metrics are reported as the mean over ten independent runs with different random seeds.

\subsection{Experimental Results}
\subsubsection{Ablation Study}
\label{ablation_study}

\begin{figure}[t]
    \centering
    \begin{subfigure}{\linewidth}
        \centering
        \includegraphics[width=\linewidth]{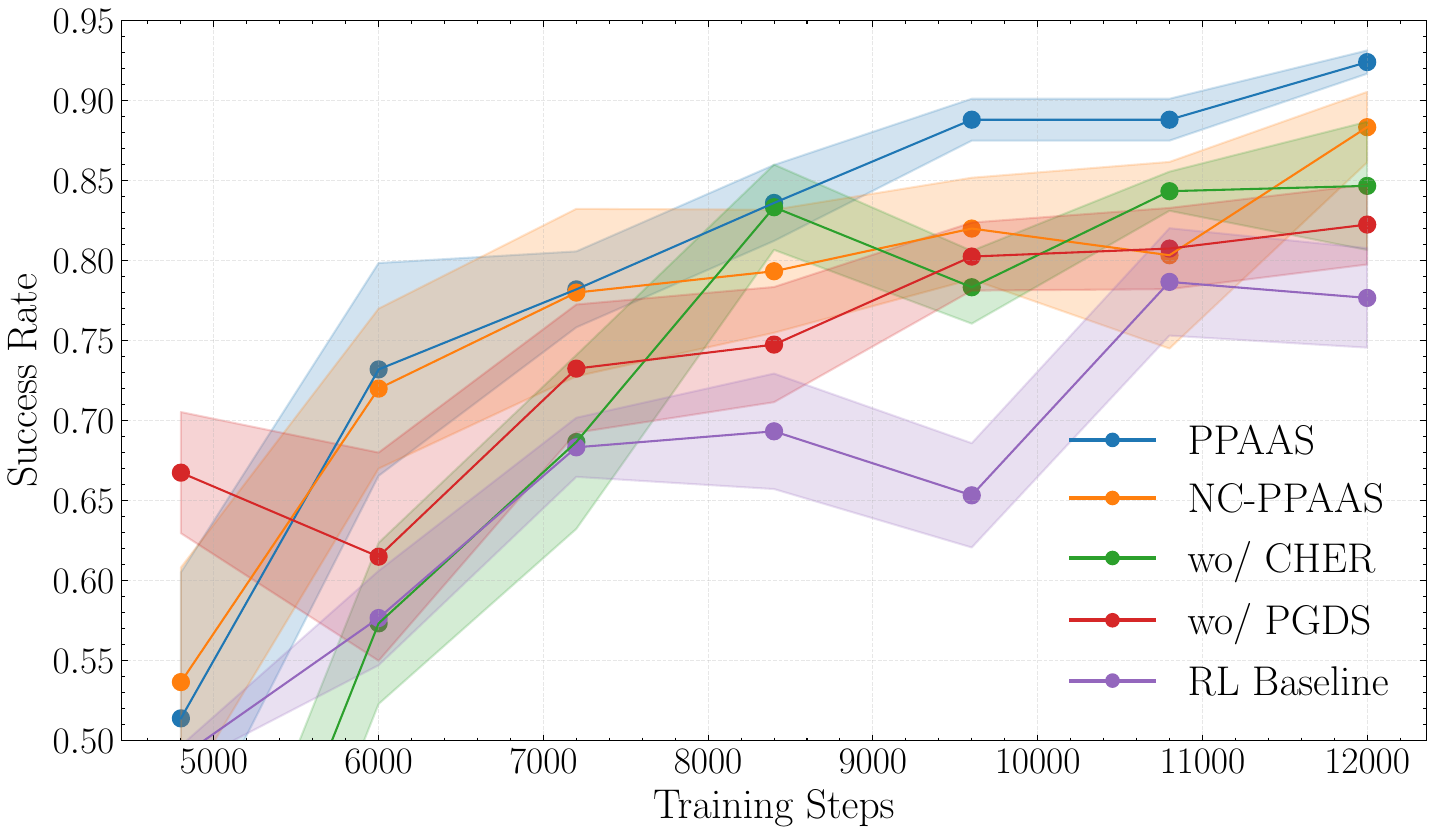}
        \caption{Success rate progression during training. The success rate is reported every 1200 steps and omitted for the initial 4800 training steps.}
        \label{fig: ablation study}
    \end{subfigure}

    \vspace{5pt}  

    \begin{subfigure}{\linewidth}
        \centering
        \includegraphics[width=\linewidth]{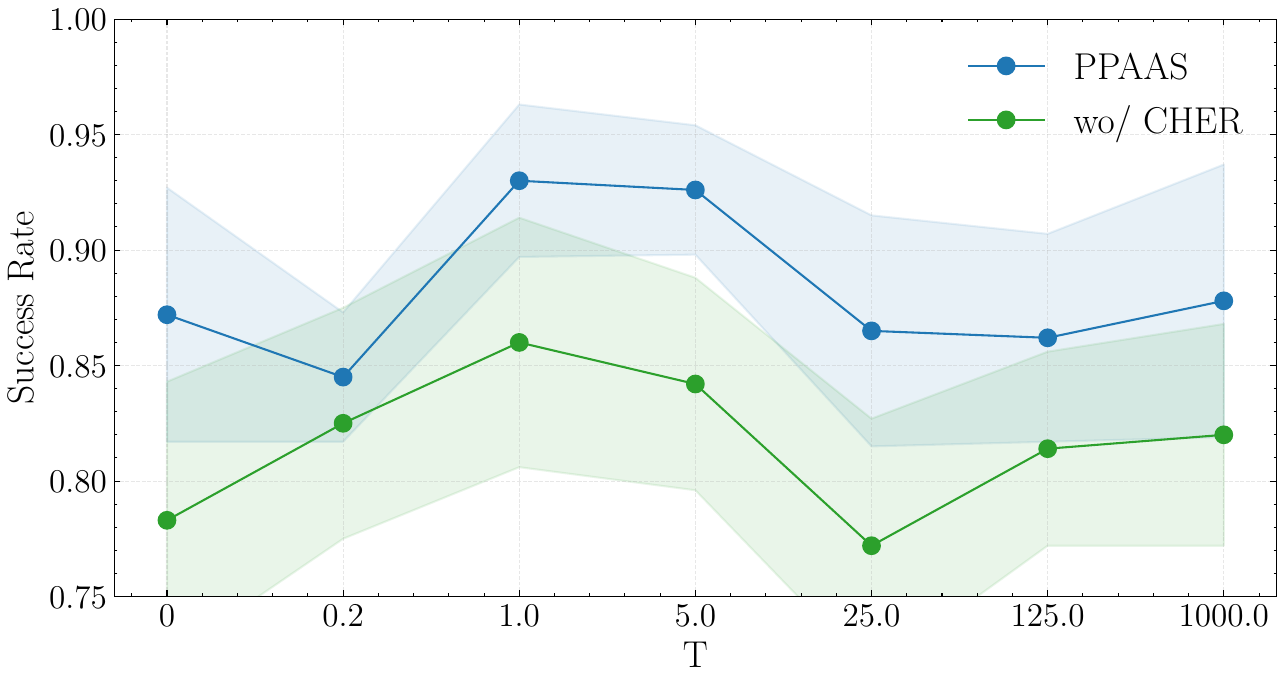}
        \caption{Success rate for different $T$ values; $T{=}0$ denotes greedy sampling.}
        \label{fig: ablation study temp}
    \end{subfigure}

    \caption{
        Ablation study on the TSA benchmark. The shaded region indicates the standard deviation across random seeds.
    }
    \label{fig:ablation_combined}
    \vspace{-10pt}
\end{figure}

Our experiments in Fig.~\ref{fig: ablation study} include an ablation study on the TSA benchmark to evaluate the effectiveness of the components introduced in our method. All variants preserve the underlying GCRL setup, including the SoF simulation framework and a reward formulation with $\alpha\!=\!0$.

Fig.~\ref{fig: ablation study} shows that \texttt{PPAAS}, which integrates both PGDS and CHER, consistently outperforms its ablated variants from step 6000 onward.
Also, it demonstrates that both PGDS and CHER contribute significantly to success rate improvement. Omitting either component results in an success rate drop exceeding 8\% by the end of training, highlighting their individual importance.
We also examine a non-conservative variant \texttt{NC-PPAAS}, where the virtual rewards for relabeled goals are not recomputed conservatively ($R'\!=\! R$).
Although this variant achieves a comparable success rate to the conservative version, it exhibits higher variance across random seeds.
This suggests that conservative reward computation enhances robustness, as evidenced by the lower standard deviation.
We also note that the RL baseline using only the SoF framework fails to achieve high success rate.
This is likely due to degraded sample quality, as skipping full-corner simulations removes critical supervision signals and thus reduces guidance for the agent.


    

We then demonstrate the effectiveness of sampling challenging goals from the set of non-\textit{Pareto-dominated} candidates with estimated Q-values in Fig.~\ref{fig: ablation study temp}. The results show that both uniform and greedy Q-value-based goal sampling yield lower success rates and greater instability, as indicated by larger deviations across seeds.

\subsubsection{Comparison with Related Work}
\label{comparison with SOTA}
In this section, we compare the performance of our method to prior works~\cite{cao2024roseoptrobustefficientanalog,shi2022robustanalogfastvariationawareanalog,settaluri2020autocktdeepreinforcementlearning,bo_analog} with the aforementioned three metrics.
For evaluation, we report the success rate using the best-performing model checkpoint obtained during training.
We also include results for the RL baseline from Section \ref{ablation_study}, as well as a variant of our method that incorporates a PVT consistency term with $\alpha=10$ during training.

Since~\cite{shi2022robustanalogfastvariationawareanalog,bo_analog} adopt a single-task approach rather than a multi-goal framework, we set their target specifications to the most stringent design point within the given range (e.g., (Delay, Switching Power) = (0.1\,ns, 0.05\,nW) for the comparator).  
The success rate is then computed by checking whether the achieved specifications—corresponding to the design parameters that yield the highest reward during training—meet or exceed each evaluation target goal.

Because~\cite{autockt,bo_analog} that do not account for PVT variations trivially fail under PVT variations, we adjust their configurations for a fair comparison.
Although all simulations are conducted only in the nominal corner, we expand the training goal range by 20\% for~\cite{autockt} and condition the actor on 20\% harder goals during deployment.
For~\cite{bo_analog}, we increase each target specification by 20\% during training.

As shown in Table~\ref{tab:performance_metrics}, \name consistently achieves the highest success rate and simulation efficiency on the TSO, CMA, and LDO benchmarks. For the COMP benchmark, it achieves the highest simulation efficiency.
Specifically, our method improves sample efficiency (success rate) by $\sim$1.6$\times$ and simulation efficiency by $\sim$4.1$\times$, averaged across the benchmarks, compared to prior methods.
The improvements in Table~\ref{tab:performance_metrics} are computed by taking the ratio between the best score achieved by \texttt{PPAAS} or its variant with $\alpha\!=\!10$ and the best score reported by prior works for SR and S$_\text{sim}$. For S$_\text{dev}$, where lower values indicate better performance, the ratio is inverted to maintain consistency in the interpretation of improvement.

Although not shown in Table~\ref{tab:performance_metrics}, we observe that \name never fails to reach the region of interest (i.e., the success rate is never zero), demonstrating strong robustness.
In contrast, RoSE-Opt~\cite{cao2024roseoptrobustefficientanalog} frequently fails to find feasible solutions for the CMA benchmark.
We also find that the pruning strategy based on K-means clustering of PVT corners~\cite{shi2022robustanalogfastvariationawareanalog} often fails to reach the region of interest under varying goals.
While it achieves the highest success rate in the COMP benchmark, it underperforms in multi-goal settings where the critical PVT corners vary across goals.
This suggests that clustering-based pruning is only effective when the number of specifications is small and corner sensitivity is consistent across goals.

Furthermore, we observe that incorporating PGDS does not improve the success rate for the COMP benchmark compared to the baseline, although it reduces the \#sim.
This is likely due to the low dimensionality of the target goals—only two specifications—which causes PGDS to prematurely suggest overly difficult goals once the goal buffer becomes sufficiently populated.
In such cases, the curriculum progression becomes unnecessarily aggressive, leading to degraded performance.

Finally, we find that including the PVT consistency term during training yields the lowest S$_\text{dev}$ among all methods, indicating a more robust parameter setting.
However, this comes at the expense of a reduced success rate for the TSA, CMA, and COMP benchmarks, highlighting a trade-off between robustness and goal reachability.
An exception is observed in the LDO benchmark, where incorporating the PVT consistency term actually improves the success rate.
This suggests that the consistency term can, in fact, enhance generalizability when the optimization direction aligns well with the primary objective.
Note that S$_\text{dev}$ is calculated only when an episode succeeds in meeting the specifications; thus, cases with zero success rate do not contribute to the statistic.

\subsubsection{Runtime Analysis}
\begin{figure}
\includegraphics[width=1.0\linewidth]{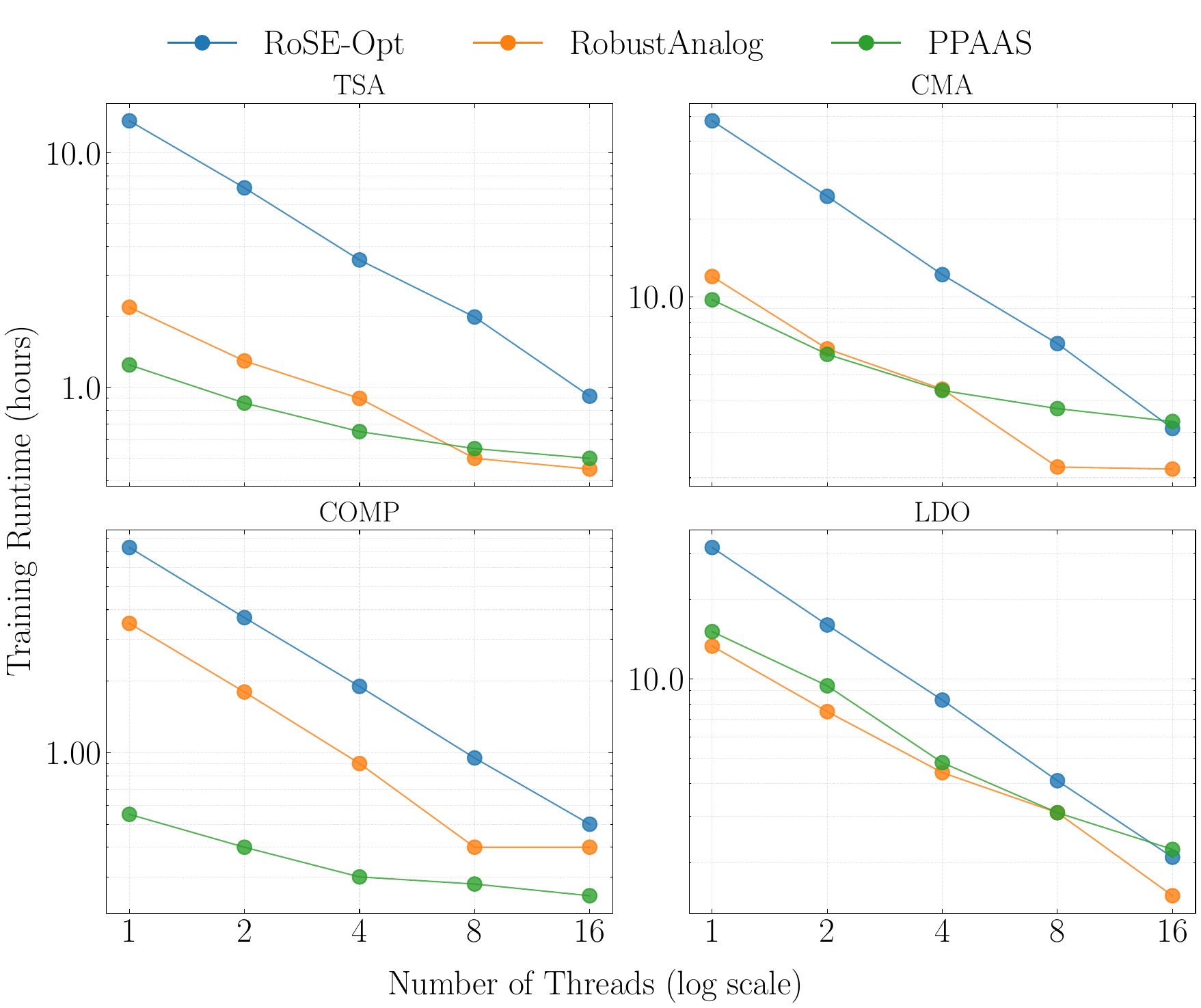}
    \caption{
    Training runtime for each benchmark using 1, 2, 4, 8, and 16 multiprocessing units.
    The runtime is independent of the SR.
    }
    \label{fig: runtime}
    \vspace{-10pt}
\end{figure}
To further demonstrate the efficiency of our training procedure, we measure the wall-clock runtime under varying computational resources, irrespective of the agent's training quality, and compared with prior works that consider PVT variations~\cite{cao2024roseoptrobustefficientanalog, shi2022robustanalogfastvariationawareanalog}.
Fig.~\ref{fig: runtime} shows the overall training time required to train the agent while varying the number of multiprocessing units.  
While full-corner simulations are counted as $N$ individual simulations when computing the \#sim, the corresponding runtime does not scale linearly since we can parallelize the simulations across PVT corners.
The single circuit simulation time per PVT corner is approximately 130ms, 320ms, 100ms, and 450ms for the TSA, CMA, COMP, and LDO benchmarks, respectively. Considering that the final goal buffer sizes were 140, 120, 8, and 700, respectively, PDGS sampling neither impedes training nor incurs significant memory overhead, with goal sampling taking less than 10ms per episode.
While runtime differences are minimal with 16 multiprocessing units, RoSE-Opt shows substantial increases as resources decrease, though RobustAnalog remains faster due to its corner pruning strategy.
The number of pruned corners for RobustAnalog is 5, 6, 8, and 9, respectively.
In contrast, our method exhibits only a marginal increase in runtime, highlighting its efficiency in resource-constrained environments.
An exception is the LDO benchmark, where nominal and full-corner simulations occur in roughly equal proportion due to the already well-optimized initial state $\mathbf{s}_0$ from~\cite{sky130_ldo_rl}, thereby diminishing the runtime reduction typically achieved by the SoF strategy.

\section{Conclusion}
\label{conclusion}
In this paper, we proposed \texttt{PPAAS}, a PVT and Pareto Aware Analog Sizing framework based on goal-conditioned reinforcement learning. Built upon a hierarchical Skip-on-Fail simulation setup, \name enhances efficiency and robustness in multi-corner environments while maintaining strong generalization capability. The framework incorporates Pareto-Dominant Goal Sampling, which constructs an automatic curriculum by selecting non-trivial goals, and Conservative Hindsight Experience Replay, which assigns conservative virtual rewards to support stable policy learning. Experimental results on diverse analog benchmarks demonstrate the effectiveness of \name in improving sample and simulation efficiency, establishing it as a practical solution for PVT-aware analog design automation in resource-constrained environments.

\section*{Acknowledgment}
This work is supported in part by NSF CCF-2112665, Samsung, SRC, Natcast AIDRFIC Program, and an equipment donation from Nvidia.

\newpage
\clearpage
{
\scriptsize
\small
\footnotesize
\bibliographystyle{IEEEtran}
\bibliography{./ref/Top_sim, ./ref/general, ./ref/PD.bib, ./ref/analog.bib, ./ref/analog_yield.bib,./ref/ML.bib, ./ref/packaging.bib}

\begin{thebibliography}{10}
\providecommand{\url}[1]{#1}
\csname url@samestyle\endcsname
\providecommand{\newblock}{\relax}
\providecommand{\bibinfo}[2]{#2}
\providecommand{\BIBentrySTDinterwordspacing}{\spaceskip=0pt\relax}
\providecommand{\BIBentryALTinterwordstretchfactor}{4}
\providecommand{\BIBentryALTinterwordspacing}{\spaceskip=\fontdimen2\font plus
\BIBentryALTinterwordstretchfactor\fontdimen3\font minus \fontdimen4\font\relax}
\providecommand{\BIBforeignlanguage}[2]{{%
\expandafter\ifx\csname l@#1\endcsname\relax
\typeout{** WARNING: IEEEtran.bst: No hyphenation pattern has been}%
\typeout{** loaded for the language `#1'. Using the pattern for}%
\typeout{** the default language instead.}%
\else
\language=\csname l@#1\endcsname
\fi
#2}}
\providecommand{\BIBdecl}{\relax}
\BIBdecl

\bibitem{Analog_TCAS_GA}
G.~Wolfe and R.~Vemuri, ``{Extraction and Use of Neural Network Models in Automated Synthesis of Operational Amplifiers},'' \emph{IEEE TCAS I}, 2003.

\bibitem{bo_analog}
W.~Lyu, F.~Yang, C.~Yan, D.~Zhou, and X.~Zeng, ``{Batch Bayesian Optimization via Multi-objective Acquisition Ensemble for Automated Analog Circuit Design},'' in \emph{Proc.~ICML}.\hskip 1em plus 0.5em minus 0.4em\relax PMLR, 2018, pp. 3306--3314.

\bibitem{complex_circuit}
J.~Zhang, J.~Bao, Z.~Huang, X.~Zeng, and Y.~Lu, ``{Automated Design of Complex Analog Circuits with Multiagent Based Reinforcement Learning},'' in \emph{Proc.~DAC}, 2023.

\bibitem{complex_circuit2}
H.~Sun, Z.~Bi, W.~Jiang, Y.~Lu, C.~Yan, F.~Yang, W.~Hu, S.-G. Wang, D.~Zhou, and X.~Zeng, ``{EVDMARL: Efficient Value Decomposition-Based Multi-Agent Reinforcement Learning with Domain-Randomization for Complex Analog Circuit Design Migration},'' in \emph{Proc.~DAC}, 2024.

\bibitem{parastic}
M.~Liu, W.~J. Turner, G.~F. Kokai, B.~Khailany, D.~Z. Pan, and H.~Ren, ``{Parasitic-Aware Analog Circuit Sizing with Graph Neural Networks and Bayesian Optimization},'' in \emph{Proc.~DATE}, 2021.

\bibitem{cronus}
Y.~Oh, D.~Kim, Y.~Lee, and B.~Hwang, ``{CRONuS: Circuit Rapid Optimization with Neural Simulator},'' in \emph{Proc.~DATE}, 2024.

\bibitem{pvtsizing}
Z.~Kong, X.~Tang, W.~Shi, Y.~Du, Y.~Lin, and Y.~Wang, ``{PVTSizing: A TuRBO-RL-Based Batch-Sampling Optimization Framework for PVT-Robust Analog Circuit Synthesis},'' in \emph{Proc.~DAC}, 2024.

\bibitem{Yang_2021}
K.-E. Yang, C.-Y. Tsai, H.-H. Shen, C.-F. Chiang, F.-M. Tsai, C.-A. Wang, Y.~Ting, C.-S. Yeh, and C.-T. Lai, ``{Trust-Region Method with Deep Reinforcement Learning in Analog Design Space Exploration},'' in \emph{Proc.~DAC}, 2021.

\bibitem{shi2022robustanalogfastvariationawareanalog}
W.~Shi, H.~Wang, J.~Gu, M.~Liu, D.~Pan, S.~Han, and N.~Sun, ``{RobustAnalog: Fast Variation-Aware Analog Circuit Design Via Multi-task RL},'' \emph{arXiv preprint arXiv:2207.06412}, 2022.

\bibitem{cao2024roseoptrobustefficientanalog}
W.~Cao, J.~Gao, T.~Ma, R.~Ma, M.~Benosman, and X.~Zhang, ``{RoSE-Opt: Robust and Efficient Analog Circuit Parameter Optimization with Knowledge-infused Reinforcement Learning},'' \emph{arXiv preprint arXiv:2407.19150}, 2024.

\bibitem{Analog_DAC20_Wang}
H.~Wang, K.~Wang, J.~Yang, L.~Shen, N.~Sun, H.-S. Lee, and S.~Han, ``{GCN-RL Circuit Designer: Transferable Transistor Sizing with Graph Neural Networks and Reinforcement Learning},'' in \emph{Proc.~DAC}, 2020.

\bibitem{DNNopt_Budak}
A.~F. Budak, P.~Bhansali, B.~Liu, N.~Sun, D.~Z. Pan, and C.~V. Kashyap, ``{DNN-Opt: An RL Inspired Optimization for Analog Circuit Sizing Using Deep Neural Networks},'' in \emph{Proc.~DAC}, 2021, pp. 1219--1224.

\bibitem{gmid}
M.~Choi, Y.~Choi, K.~Lee, and S.~Kang, ``{Reinforcement Learning-based Analog Circuit Optimizer Using gm/ID for Sizing},'' in \emph{Proc.~DAC}, 2023.

\bibitem{autockt}
K.~Settaluri, Z.~Liu, R.~Khurana, A.~Mirhaj, R.~Jain, and B.~Nikolic, ``{Automated Design of Analog Circuits Using Reinforcement Learning},'' \emph{IEEE TCAD}, 2022.

\bibitem{ppo}
J.~Schulman, F.~Wolski, P.~Dhariwal, A.~Radford, and O.~Klimov, ``{Proximal Policy Optimization Algorithms},'' \emph{arXiv preprint arXiv:1707.06347}, 2017.

\bibitem{circuit_pvt_1}
S.~Natarajan, M.~Breuer, and S.~Gupta, ``{Process Variations and Their Impact on Circuit Operation},'' in \emph{Proceedings 1998 IEEE International Symposium on Defect and Fault Tolerance in VLSI Systems}, 1998.

\bibitem{circuit_pvt_2}
I.~Ghorbel, F.~Haddad, W.~Rahajandraibe, and M.~Loulou, ``{A Subthreshold Low-Power CMOS LC-VCO with High Immunity to PVT Variations},'' \emph{Analog Integr. Circuits Signal Process.}, 2017.

\bibitem{Schaul2015UniversalVF}
T.~Schaul, D.~Horgan, K.~Gregor, and D.~Silver, ``{Universal Value Function Approximators},'' in \emph{Proc.~ICML}, 2015.

\bibitem{andrychowicz2018hindsightexperiencereplay}
M.~Andrychowicz, F.~Wolski, A.~Ray, J.~Schneider, R.~Fong, P.~Welinder, B.~McGrew, J.~Tobin, P.~Abbeel, and W.~Zaremba, ``{Hindsight Experience Replay},'' in \emph{Proc.~NIPS}, 2018.

\bibitem{settaluri2020autocktdeepreinforcementlearning}
K.~Settaluri, A.~Haj-Ali, Q.~Huang, K.~Hakhamaneshi, and B.~Nikolic, ``{AutoCkt: Deep Reinforcement Learning of Analog Circuit Designs},'' in \emph{Proc.~DATE}, 2020.

\bibitem{curriculum_learning}
Y.~Bengio, J.~Louradour, R.~Collobert, and J.~Weston, ``Curriculum learning,'' in \emph{Proc.~ICML}, 2009.

\bibitem{automatic_curriculum}
Y.~Zhang, P.~Abbeel, and L.~Pinto, ``Automatic curriculum learning through value disagreement,'' in \emph{Proc.~NIPS}, 2020.

\bibitem{quasi_reward}
K.~Valieva and B.~Banerjee, ``Quasimetric value functions with dense rewards,'' \emph{arXiv preprint arXiv:2409.08724}, 2024.

\bibitem{sac}
T.~Haarnoja, A.~Zhou, P.~Abbeel, and S.~Levine, ``{Soft Actor-Critic: Off-Policy Maximum Entropy Deep Reinforcement Learning with a Stochastic Actor},'' in \emph{Proc.~ICML}.\hskip 1em plus 0.5em minus 0.4em\relax PMLR, 2018, pp. 1861--1870.

\bibitem{ddpg}
T.~P. Lillicrap, J.~J. Hunt, A.~Pritzel, N.~Heess, T.~Erez, Y.~Tassa, D.~Silver, and D.~Wierstra, ``Continuous control with deep reinforcement learning,'' \emph{arXiv preprint:1509.02971}, 2019.

\bibitem{pessimistic_reward}
H.~Li, X.-H. Zhou, X.-L. Xie, S.-Q. Liu, Z.-Q. Feng, X.-Y. Liu, M.-J. Gui, T.-Y. Xiang, D.-X. Huang, B.-X. Yao, and Z.-G. Hou, ``{CROP: Conservative Reward for Model-based Offline Policy Optimization},'' \emph{arXiv preprint arXiv:2310.17245}, 2023.

\bibitem{analoggym}
J.~Li, H.~Zhi, R.~Lyu, W.~Li, Z.~Bi, K.~Zhu, Y.~Zeng, W.~Shan, C.~Yan, F.~Yang, Y.~Li, and X.~Zeng, ``{AnalogGym: An Open and Practical Testing Suite for Analog Circuit Synthesis},'' \emph{arXiv preprint arXiv:2409.08534}, 2024.

\bibitem{fmc_3s_amp}
M.~Tan and W.-H. Ki, ``{A Cascode Miller-Compensated Three-Stage Amplifier With Local Impedance Attenuation for Optimized Complex-Pole Control},'' \emph{IEEE Journal Solid-State Circuits}, vol.~50, no.~2, pp. 440--449, 2015.

\bibitem{sky130_ldo_rl}
Z.~Li and A.~C. Carusone, ``{Design and Optimization of Low-Dropout Voltage Regulator Using Relational Graph Neural Network and Reinforcement Learning in Open-Source SKY130 Process},'' in \emph{Proc.~ICCAD}, 2023.

\bibitem{gf180mcu_pdk}
\BIBentryALTinterwordspacing
\emph{{GF180MCU PDK}}, GlobalFoundry. [Online]. Available: \url{https://github.com/google/gf180mcu-pdk}
\BIBentrySTDinterwordspacing

\end{thebibliography}
}

\end{document}